# Spin-orbit coupling driven insulating state in hexagonal iridates $Sr_3MIrO_6$ (M = Sr, Na and Li)

Xing Ming,[1*] Xiangang Wan,[2] Carmine Autieri,[3,4] Jianfeng Wen,[1] Xiaojun Zheng[1]

*1. College of Science, Guilin University of Technology, Guilin 541004, People's Republic of China*

*2. National Laboratory of Solid State Microstructures, Collaborative Innovation Center of Advanced Microstructures, College of Physics, Nanjing University, Nanjing 210093, People's Republic of China*

*3. Consiglio Nazionale delle Ricerche CNR-SPIN, UOS L'Aquila, Sede Temporanea di Chieti, 66100 Chieti, Italy*

*4. International Research Centre Magtop, Polish Academy of Sciences, Aleja Lotnik ów 32/46, PL-02668 Warsaw*

**ABSTRACT**

The spin-orbit coupling (SOC) interactions, electron correlation effects and Hund coupling cooperate and compete with each other, leading to novel properties, quantum phase and non-trivial topological electronic behavior in iridium oxides. Because of the well separated $IrO_6$ octahedra approaching cubic crystal-field limit, the hexagonal iridates $Sr_3MIrO_6$ (M = Sr, Na and Li) serves as a canonical model system to investigate the underlying physical properties that arises from the novel $J_{eff}$ state. Based on density functional theory calculations complemented by Green's function methods, we systematically explore the critical role of SOC on the electronic structure and magnetic properties of $Sr_3MIrO_6$. The crystal-field splitting combined with correlation effects are insufficient to account for the insulating nature, but the SOC interactions is the intrinsic source to trigger the insulating ground states in these hexagonal iridates. The decreasing geometry connectivity of $IrO_6$ octahedra gives rise to the increasing of effective electronic correlations and SOC interactions, tuning the hexagonal iridates from low-spin $J_{eff} = 1/2$ states with large local magnetic moments for the $Ir^{4+}$ ($5d^5$) ions in $Sr_4IrO_6$ to nonmagnetic singlet $J_{eff} = 0$ states without magnetic moments for the $Ir^{5+}$ ($5d^4$) ions in $Sr_3NaIrO_6$ and $Sr_3LiIrO_6$. The theoretical calculated results are in good agreement with available experimental data, and explain the magnetic properties of $Sr_3MIrO_6$ well.

*Corresponding author: mingxing@glut.edu.cn

# I. INTRODUCTION

In the latest decade, iridium oxides (iridates) have attracted growing attention, due to the fact that these materials host a number of intriguing phenomena and novel physical properties [1, 2, 3, 4, 5]. Special interest in the iridates has been stimulated by the pioneering work of Kim *et al*, where a novel spin-orbit induced Mott insulating state has been discovered in $Sr_2IrO_4$ [6, 7]. They proposed that the large octahedral crystal-field splitting between the triply degenerate $t_{2g}$ states and doubly degenerate $e_g$ states, together with the strong spin-orbit coupling (SOC) interactions generate the quartet $j_{eff}$ = 3/2 and doublet $j_{eff}$ = 1/2 states [1, 6]. The electron correlation effects cooperate with the SOC interactions, creating a novel $J_{eff}$ = 1/2 insulating state in $Sr_2IrO_4$. The $J_{eff}$ = 1/2 state has been proposed to be a common ingredient in iridates, which renewed the attentions in the interplay of the electron correlations, geometry connectivity and SOC interactions [1-5, 8, 9, 10, 11, 12, 13]. Depending on the relative strength of on-site Coulomb repulsion, geometry connectivity and SOC interactions, the iridates have been proposed as promising candidates for exotic phases [1, 4], such as topological insulators [4, 14], spin-orbit coupled Mott insulator [6, 7], giant magnetic anisotropy [10, 11], superconductors [15, 16, 17, 18], Weyl semimetals [19, 20], Heavy-mass magnetic modes [21], spin liquids and spin ices [22, 23].

The lattice degree of freedom plays a critical role in iridates, giving rise to abundant structures in iridates, like the Ruddlesden-Popper series $Sr_{n+1}Ir_nO_{3n+1}$ (n = 1, 2 and ∞), pyrochlore $R_2Ir_2O_7$ (R is the rare-earth element or Y), two-dimensional honeycomb geometry, hexagonal perovskite, post-perovskite and double-perovskite structure [3, 5]. In the present paper, we focus on three hexagonal iridates, $Sr_3MIrO_6$ (M = Sr, Na and Li) [24, 25, 26, 27, 28]. These $Sr_3MIrO_6$ iridates crystallize in the rhombohedral $K_4CdCl_6$ structure (space group $R\bar{3}c$) with six formula units (f. u.) in the hexagonal lattice (**Fig. 1** (a)), whereas with two f. u. per rhombohedral primitive cell (**Fig. 1** (b)). According to the structural symmetry, Sr, M, Ir, and O atoms occupy four nonequivalent crystallographic sites: 18*e* (x, 0, 0.25), 6*a* (0, 0, 0.25), 6*b* (0, 0, 0) and 36*f* (x, y, z) sites, respectively [24, 25, 26, 27, 28]. The hexagonal structure is often viewed as chains parallel to the *c* axis, which constructed from face-sharing $IrO_6$ octahedra with trigonal prismatically coordinated $MO_6$ polyhedra [29]. $Sr_3MIrO_6$ are isostructural with $Sr_4PtO_6$ [24, 25], $Ca_4IrO_6$ [30, 31, 32], $Sr_4RhO_6$ [33], $Sr_3CoIrO_6$ and $Sr_3NiIrO_6$ [34, 35]. $Sr_4IrO_6$ was first prepared by Randall and Katz [24], which was belong to one of the stable compounds in the Sr-Ir-O system under atmospheric conditions [36, 37, 38, 39]. Magnetic susceptibility measurements revealed the onset of antiferromagnetic (AFM) ordering below 12 K, and no evidence for weak ferromagnetism in $Sr_4IrO_6$ [40]. Substituting one quarter of the $Sr^{2+}$ ions with $Li^+$ or $Na^+$ ions results in the Ir ions transform from a formal tetravalent ($Ir^{4+}$, $5d^5$) to pentavalent ($Ir^{5+}$, $5d^4$) [26, 27, 28]. These iridates containing $Ir^{5+}$ ions show temperature-independent paramagnetic character in $Sr_3NaIrO_6$ and $Sr_3LiIrO_6$ [27, 28].

Relative to other extensively investigated iridates with $5d^5$ electronic configurations, far less attention have been paid to the physical properties of hexagonal iridates with $5d^4$ electronic configurations. In the strong SOC limit, the $J_{eff} = 0$ state has been proposed to explain the absence of magnetic ordering in the pentavalent ($Ir^{5+}$, $5d^4$) iridates with the four electrons fully occupying the lower $j_{eff} = 3/2$ quadruplet [41]. On the other hand, when the Hund coupling ($J_H$) dominate over SOC interactions, a low-spin $S = 1$ state is realized in iridates with $5d^4$ electronic configurations [3]. Moreover, crystal-field effects and non-cubic structural distortions of the octahedral coordination-environment, as well as the geometry connectivity of $IrO_6$ octahedra further influence the electronic correlations and SOC interactions [4, 8, 9]. As a result, even possessing strong enough SOC, so far an absolutely nonmagnetic (NM) state has been seldom realized in the $5d^4$ systems. Experimental and theoretical results shown hexagonal $Ca_4IrO_6$ residing in the cubic crystal-field limit required for a canonical unmixed $J_{eff} = 1/2$ state [32]. By comparison, the $IrO_6$ octahedra in $Sr_3MIrO_6$ (especially for M = Sr and Na) reside much more close to the ideal cubic crystal-field limit [25, 26, 27]. As shown in **Fig. 1**, the $IrO_6$ octahedra are well separated with each other and disconnected in these hexagonal $Sr_3MIrO_6$. Inside the octahedra, the Ir-O bond lengths are identical, accompanying with tiny deviations of O-Ir-O bond angles from 90° [25, 26, 27, 40, 42]. These hexagonal $Sr_3MIrO_6$ iridates provide almost ideal cubic crystal-field, serving as a good platform to explore the cooperation and competing of the comparable energy scales of SOC interactions, electron correlation and Hund coupling. In addition, first-principles electronic structure calculation is an ideal tool to provide insight into the underlying role of SOC on the electronic structure and magnetic properties of the title iridates $Sr_3MIrO_6$.

Based on density functional theory (DFT) first-principles electronic structure calculations complemented by Green's function method, we reveal the impact of the local crystal structure, the connectivity of IrO$_6$ octahedra and SOC interactions on the electronic structure of these iridates. The SOC interactions is the key determinant factor in opening the insulating band gaps in Sr$_3$MIrO$_6$, rather than the Coulomb interaction. Sr$_4$IrO$_6$ show large local magnetic moments with unmixed $J_{eff}$ = 1/2 characters, whereas the strong SOC limit leads to NM singlet $J_{eff}$ = 0 states in Sr$_3$MIrO$_6$ (M = Na and Li). Our theoretical results supply a meaningful complement to the hectic field of iridates with strong SOC interactions. The remainder of the paper is organized in the following way. In Sec. II, we provide the details of our computational techniques and describe the crystal structure of Sr$_3$MIrO$_6$. Section III is devoted to calculated results and discussions followed by the conclusions in Sec. IV.

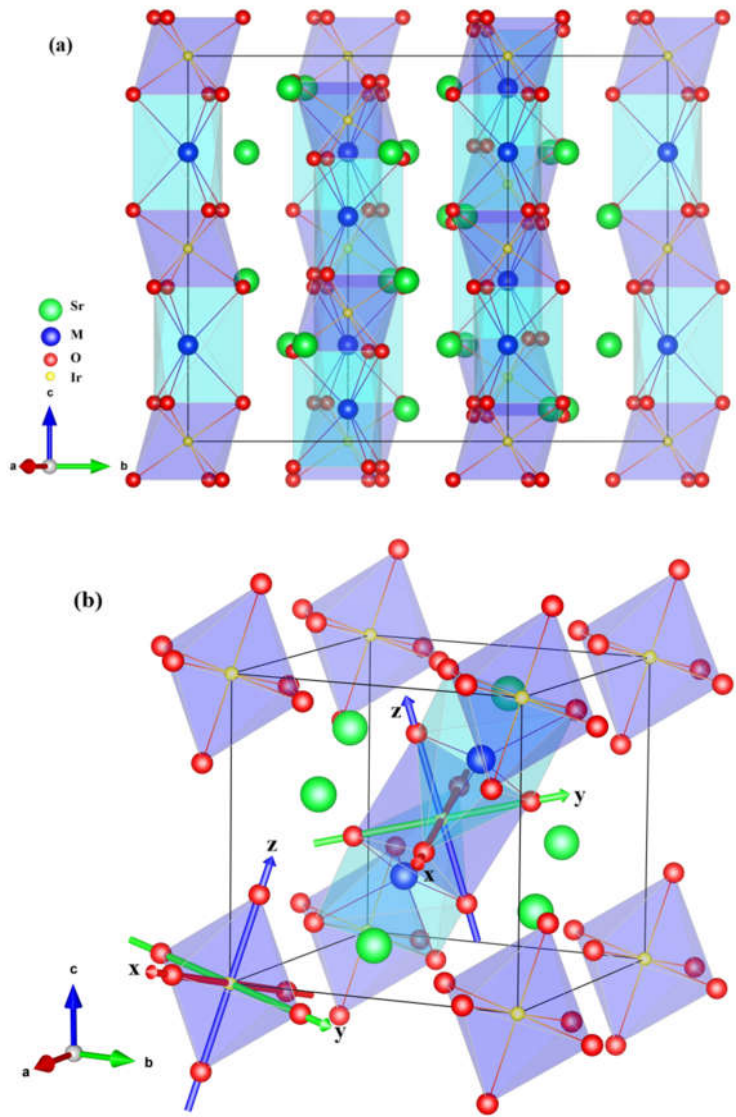

**Fig. 1** (a) Hexagonal crystal structure and (b) Rhombohedral primitive unit cell of Sr$_3$MIrO$_6$ (M =

Sr, Na and Li). For better clarity, local coordinate system (*x, y, z*) is defined for the rhombohedral primitive unit cell, with *x, y, z* approximately along one of the O-Ir-O bond directions in the IrO$_6$ octahedron.

## II. COMPUTATIONAL DETAILS AND CRYSTAL STRUCTURE

The first-principles DFT calculations have been carried out with the projector augmented wave (PAW) method [43, 44] as implemented in the Vienna *ab initio* simulation package (VASP) code [45], together with the generalized gradient approximation (GGA) [46]. The rotationally invariant DFT + *U* method introduced by Liechtenstein *et al*. is employed to consider the correlations effects [47]. All the calculations are performed with onsite-Coulomb interactions *U* = 2 eV and the Hund's coupling $J_H$ = 0.2 eV for the Ir atoms [8, 11, 12]. SOC has been taken into account with unconstrained noncollinear magnetism settings. The structural optimization and electronic structure calculations are performed for the rhombohedral primitive cell (2 formulas), using 7 × 7 × 7 *k*-point mesh and 520 eV cutoff-energy. The convergence threshold of the self-consistent field cycle is 10$^{-6}$ eV.

Started from the experimental lattice parameters, we optimize all atomic internal coordinates and lattice constants of the rhombohedral primitive cell. For the case of Sr$_4$IrO$_6$ in order to simulate the AFM ordering [40], we consider an antiparallel alignment of the spin moment for the two Ir ions in the rhombohedral primitive cell. For the cases of Sr$_3$MIrO$_6$ (M = Na and Li), according to the NM characteristic of the magnetic susceptibility [28], the crystal structures are relaxed by considering NM spin moment in the rhombohedral primitive cell. As presented in **TABLE I**, our theoretical calculated lattice parameters are in good agreement with available experimental data [25, 26, 27, 28], with errors less than 2% for all the lattice parameters. Different from the large tetragonal distortions in the Ruddlesden-Popper (RP) series of iridates (such as the significantly elongation in the *z* direction in Sr$_2$IrO$_4$), the IrO$_6$ octahedra are tiny distorted only in terms of O-Ir-O angle in Sr$_3$MIrO$_6$. Due to the smaller Li$^+$ ionic radius compared to those of Na$^+$ and Sr$^{2+}$, the distortion is much more pronounced in Sr$_3$LiIrO$_6$ than Sr$_3$NaIrO$_6$ and Sr$_4$IrO$_6$. Along with the decreasing of the ionic radii from Sr$^{2+}$ to Na$^+$ and finally Li$^+$, the lattice constants and the Ir-Ir bond lengths evolve smaller and smaller in Sr$_3$MIrO$_6$. The IrO$_6$ octahedra in Sr$_3$NaIrO$_6$ and Sr$_4$IrO$_6$ are much closer to the ideal cubic crystal-field limit than in Sr$_3$LiIrO$_6$ [25, 26, 27]. Our theoretical calculated results confirm that a reasonable *U* parameter and SOC have only a small influence on the crystal structure, and electronic-structure calculations are performed based on the optimized crystal structures within GGA.

**TABLE I** Theoretical calculated and experimental measured lattice constants (Å), atomic internal coordinates, Ir-O bond length (Å) and O-Ir-O bond angles (°) for $Sr_3MIrO_6$ (M = Sr, Na and Li).

| Compound | Method | Lattice constants | | Sr | O | | | Ir-O | O-Ir-O |
|---|---|---|---|---|---|---|---|---|---|
| | | $a = b$ | $c$ | x | x | y | z | | |
| $Sr_4IrO_6$ | Expt.[a] | 9.734 | 11.892 | 0.3655 | 0.1840 | 0.0268 | 0.0989 | 2.047 | 90.293 |
| | GGA[e] | 9.842 | 11.966 | 0.3658 | 0.1850 | 0.0273 | 0.0997 | 2.079 | 90.336 |
| | GGA+$U$[e] | 9.847 | 11.947 | 0.3658 | 0.1850 | 0.0273 | 0.0997 | 2.078 | 90.438 |
| | GGA+SOC[e] | 9.846 | 12.000 | 0.3664 | 0.1846 | 0.0271 | 0.1000 | 2.081 | 90.079 |
| | GGA+SOC+$U$[e] | 9.844 | 12.000 | 0.3665 | 0.1845 | 0.0270 | 0.1000 | 2.080 | 90.031 |
| $Sr_3NaIrO_6$ | Expt.[b] | 9.636 | 11.556 | 0.3574 | 0.1773 | 0.0216 | 0.0985 | 1.976 | 90.103 |
| | Expt.[d] | 9.638 | 11.585 | 0.3580 | 0.1840 | 0.0268 | 0.0989 | 2.017 | 90.906 |
| | GGA[e] | 9.712 | 11.673 | 0.3574 | 0.1798 | 0.0237 | 0.1000 | 2.016 | 90.202 |
| | GGA+$U$[e] | 9.705 | 11.664 | 0.3575 | 0.1788 | 0.0238 | 0.1000 | 2.014 | 90.142 |
| | GGA+SOC[e] | 9.732 | 11.682 | 0.3573 | 0.1796 | 0.0239 | 0.0999 | 2.016 | 90.174 |
| | GGA+SOC+$U$[e] | 9.739 | 11.679 | 0.3573 | 0.1793 | 0.0240 | 0.0998 | 2.014 | 90.137 |
| $Sr_3LiIrO_6$ | Expt.[c] | 9.636 | 11.144 | 0.3583 | 0.1745 | 0.0215 | 0.1047 | 1.971 | 91.493 |
| | Expt.[d] | 9.6419 | 11.147 | 0.3587 | 0.1840 | 0.0268 | 0.0989 | 1.993 | 92.349 |
| | GGA[e] | 9.723 | 11.218 | 0.3584 | 0.1752 | 0.0216 | 0.1064 | 2.004 | 91.878 |
| | GGA+$U$[e] | 9.717 | 11.210 | 0.3586 | 0.1752 | 0.0215 | 0.1064 | 2.002 | 91.867 |
| | GGA+SOC[e] | 9.738 | 11.234 | 0.3578 | 0.1752 | 0.0219 | 0.1062 | 2.004 | 91.822 |
| | GGA+SOC+$U$[e] | 9.743 | 11.231 | 0.3577 | 0.1752 | 0.0220 | 0.1059 | 2.002 | 91.691 |

[a]Reference [25].

[b]Reference [26].

[c]Reference [27].

[d]Reference [28].

[e]Present work.

### III. RESULTS AND DISCUSSIONS

#### A. Spin polarized electronic structure

To clarify the basic electronic structure, we firstly perform spin polarized calculations within GGA, and then consider the correlation interactions by GGA + $U$ calculations. The O 2$p$ states mainly locate at lower energy region below -1 eV (not shown here), which are separated from the higher-energy Ir 5$d$ states around Fermi level ($E_F$) by a large gap. The Ir 5$d$ orbitals have been split into the $t_{2g}$ and $e_g$ states in the octahedral crystal field. As shown in **Fig. 2**, the typical characteristics of the band structure are the isolated manifold of six $t_{2g}$ bands cross over the Fermi level, which are arising from the two Ir atoms in the rhombohedral primitive cell, and resulting in metallic electronic structure in these iridates within GGA. The $e_g$ states are fully empty and are distinctly split away from the $t_{2g}$ states by a large crystal-field splitting (not shown here). Analogous to other iridates with $Ir^{4+}$ ions, $Sr_4IrO_6$ shows low-spin 5$d^5$ ($t_{2g}^5$, $e_g^0$) electronic configurations, in line with the magnetic-susceptibility measurement results [40]. In contrast, in the cases of $Sr_3NaIrO_6$ and $Sr_3LiIrO_6$, in order to describe the real material and explore the effect of SOC successively in these iridates, we have just simulated NM states ($S = 0$) according to the NM characteristics of the magnetic-susceptibility of $Sr_3NaIrO_6$ and $Sr_3LiIrO_6$ [28].

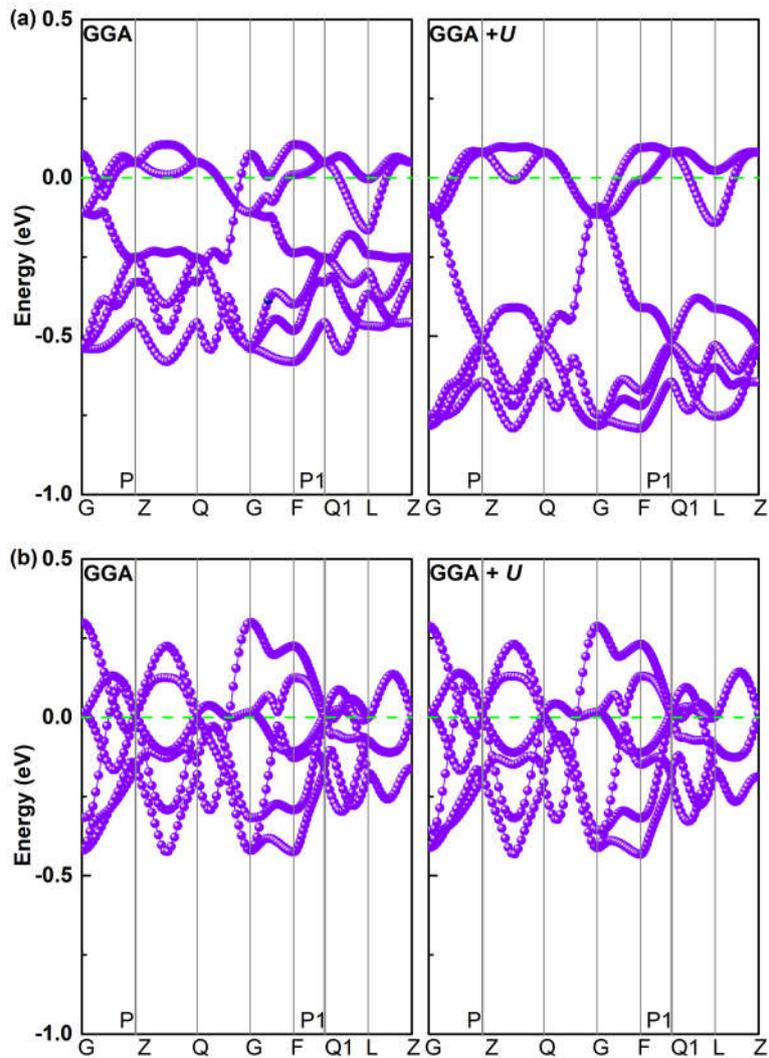

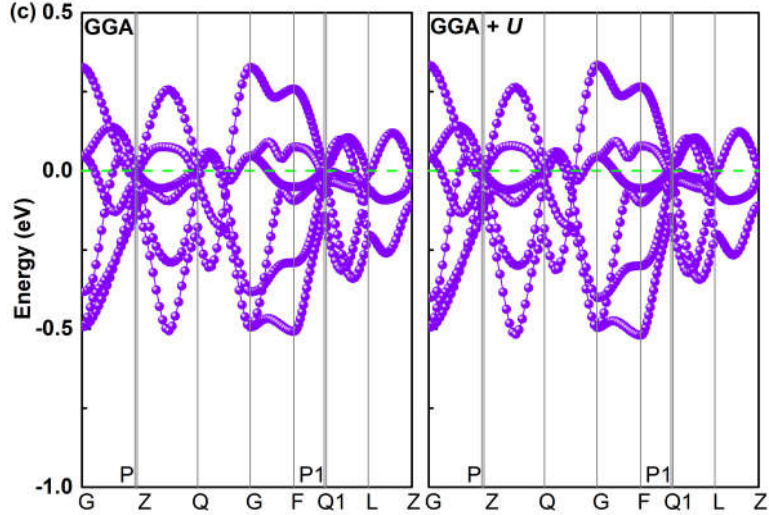

**Fig. 2** Spin polarized band structures of $Sr_3MIrO_6$ (M = Sr, Na and Li): (a) $Sr_4IrO_6$, (b) $Sr_3NaIrO_6$, (c) $Sr_3LiIrO_6$. Left and right columns are calculated by GGA and GGA + $U$, respectively. Since spin-up and spin-down states are degenerate in the AFM or NM states, only spin-up subbands are presented.

Another notable feature of the electronic structure is the bandwidth of the $t_{2g}$ states are much narrower than those of $Sr_2IrO_4$ and $SrIrO_3$ [6, 8], due to the reduced connectivity. For this reason, there are much stronger electron correlation interactions and more localized 5$d$ electronic state in $Sr_3MIrO_6$. The lower connectivity and narrower bandwidth implies more difficult hopping between the magnetic ions, is also consistent with the lower AFM ordering temperature for $Sr_4IrO_6$ (12 K) [40] relative to $Sr_2IrO_4$ (250 K) [48]. Progression from three-dimensional perovskite semimetal $SrIrO_3$ to the quasi two-dimensional layered correlated insulator $Sr_2IrO_4$, to the title compounds of $Sr_3MIrO_6$ with well separated $IrO_6$ octahedra, the ratio between electron correlations and bandwidth are increasing along with the decreasing connectivity of the $IrO_6$ octahedra [9]. Attributed to the small enough bandwidth in these $Sr_3MIrO_6$ iridates, we naturally expect that even a modest onsite Coulomb repulsion $U$ among the Ir 5$d$ states is sufficient to open the insulating gap. However, as shown in **Fig. 2**, including a reasonable Coulomb interactions parameter $U$ [8, 10, 11], the bandwidth of the $t_{2g}$ states increases a lot for $Sr_4IrO_6$ (**Fig. 2** (a)), whereas $U$ almost has no impact on the $t_{2g}$ states of $Sr_3NaIrO_6$ and $Sr_3LiIrO_6$ (**Fig. 2** (b) and (c)). The electronic structure are essentially keep the metallic character, even much larger $U$ value up to 5 eV is still unable to open the band gap in $Sr_3MIrO_6$, indicating that the crystal-field splitting combined with correlation effects are insufficient to account for the insulating nature. We'll uncover the intrinsic source to tune the insulating nature in these hexagonal iridates in the following sections.

## B. $J_{eff} = 1/2$ state of the $Sr_4IrO_6$

To gain insight into the underlying effects of SOC on the electronic structure of $Sr_4IrO_6$, we include SOC interactions by GGA + SOC calculations. As shown in **Fig. 3**, SOC significantly influences the band dispersion of the $5d$ state around Fermi level, the three-fold degenerate $t_{2g}$ states evolve into well separated $j_{eff} = 1/2$ doublet and $j_{eff} = 3/2$ quartet states with obvious gap between them, leading to an unmixed $J_{eff} = 1/2$ character. Due to the strong SOC effects and the isolated octahedra, the system reduces to a half-filed $J_{eff} = 1/2$ Hubbard system. The $j_{eff} = 1/2$ bands are further split into two pairs by a ~0.2 eV insulating gap, contributing to the valence band maximum (VBM) and the conduction band minimum (CBM).

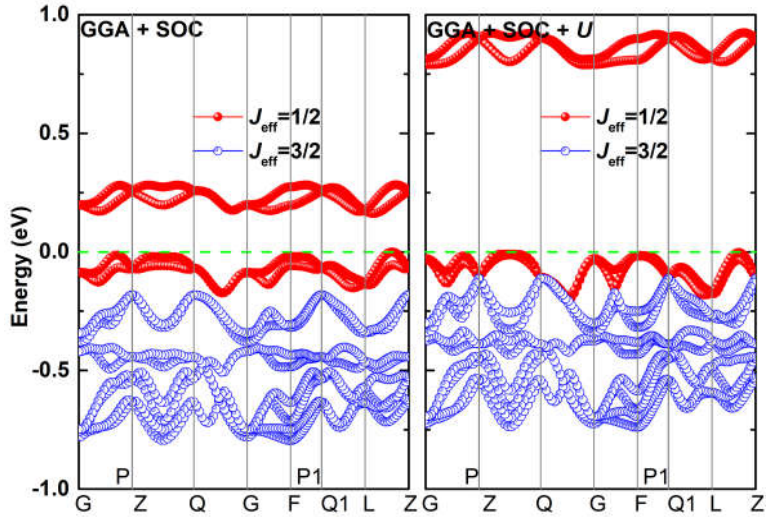

**Fig. 3** Band structures of $Sr_4IrO_6$ calculated within GGA + SOC (left) and GGA + SOC + $U$ (right). The $j_{eff} = 1/2$ doublet are shown in red while the $j_{eff} = 3/2$ quartet states are shown in blue.

Although without Coulomb interactions $U$, the $Sr_4IrO_6$ undergoes an electronic transition from the AFM metallic state to an AFM insulating state with the SOC interactions included, indicating the critical role of SOC interactions to trigger the system to be insulating. Unmixed $J_{eff} = 1/2$ magnetic insulating state also have been observed in the isostructural iridate $Ca_4IrO_6$ [32], and the fluorido-iridates molecular [49], as well as the hexafluoro iridates $Rb_2IrF_6$ [50], where also existing spatially isolated octahedra with tiny distortions. However, despite almost having perfectly undistorted octahedra, the $j_{eff} = 1/2$ and $j_{eff} = 3/2$ bands are not fully separated in $4d$ oxide, for instance in the isostructural hexagonal $Sr_4RhO_6$ [33] and the alkali metal hexafluoro rhodates

Rb$_2$RhF$_6$ [50], indicating the intrinsic SOC interactions in 4$d$ compound are weaker than those of 5$d$ iridates. On the other hand, although having large SOC effects in the first-proposed SOC induced Mott insulator Sr$_2$IrO$_4$, the $j_{\text{eff}}$ = 1/2 and $j_{\text{eff}}$ = 3/2 bands are mixed together due to obviously tetragonal structural distortions and large octahedral rotations, and it needs a modest $U$ combined with SOC to open up an insulating gap [6, 8]. These results demonstrate that the decreasing of the connectivity of the IrO$_6$ octahedra leads to an increasing of the electron-electron interaction impact [9], accompanied by an enhancement of SOC effects [4]. By introducing on-site Coulomb correlation $U$, the insulating gap further increases to ~0.8 eV in Sr$_4$IrO$_6$. The band gap in Sr$_4$IrO$_6$ is larger than that of Sr$_2$IrO$_4$ [6, 51], which is consistent with the higher activation energy and much lower electrical conductivity in Sr$_4$IrO$_6$ relative to Sr$_2$IrO$_4$ [39]. The calculated electronic structures imply that $U$ cannot play a major role on the insulating band gap, but the strong SOC interactions are essential to trigger the insulating state in Sr$_4$IrO$_6$.

### C. Magnetic property of the Sr$_4$IrO$_6$

Although magnetic susceptibility measurements shown AFM ordering below 12 K in Sr$_4$IrO$_6$, no detailed magnetic structure has been reported to date [40]. We assume the magnetic unit cell to be the same as the crystallographic hexagonal unit cell, and try to understand the possible magnetic structure by symmetry analyzing [52]. The $R\bar{3}c$ space group of Sr$_4$IrO$_6$ allows five magnetic space groups for the crystallographic unit cell: (1) $R\bar{3}c$, (2) $R\bar{3}c1'$, (3) $R\bar{3}'c$, (4) $R\bar{3}'c'$, (5) $R\bar{3}c'$, where $R\bar{3}c1'$ corresponds to the paramagnetic configuration, whereas $R\bar{3}'c$ and $R\bar{3}'c'$ are NM configurations, and the last one $R\bar{3}c'$ is ferromagnetic (FM) configuration. Therefore, $R\bar{3}c$ is the only one remaining magnetic space group can be assigned to the AFM ordering of Sr$_4$IrO$_6$ (as shown in **Fig. 4**).

**TABLE II** Relative energy (meV/f.u.), calculated spin moment ($M_S$) and orbital moment ($M_L$) (values in Bohr magnetons) for Sr$_4$IrO$_6$ within SOC + $U$.

| Configuration | Magnetic space groups | ΔE | $M_S$ | $M_L$ |
|---|---|---|---|---|
| NM | $R\bar{3}c1'$, $R\bar{3}'c$, $R\bar{3}'c'$ | 190.712 | 0 | 0 |
| FM | $R\bar{3}c'$ | 0.823 | 0.426 | 0.496 |
| AFM1 | | 0.684 | 0.420 | 0.500 |
| AFM2 | $R\bar{3}c$ | 0 | 0.417 | 0.503 |

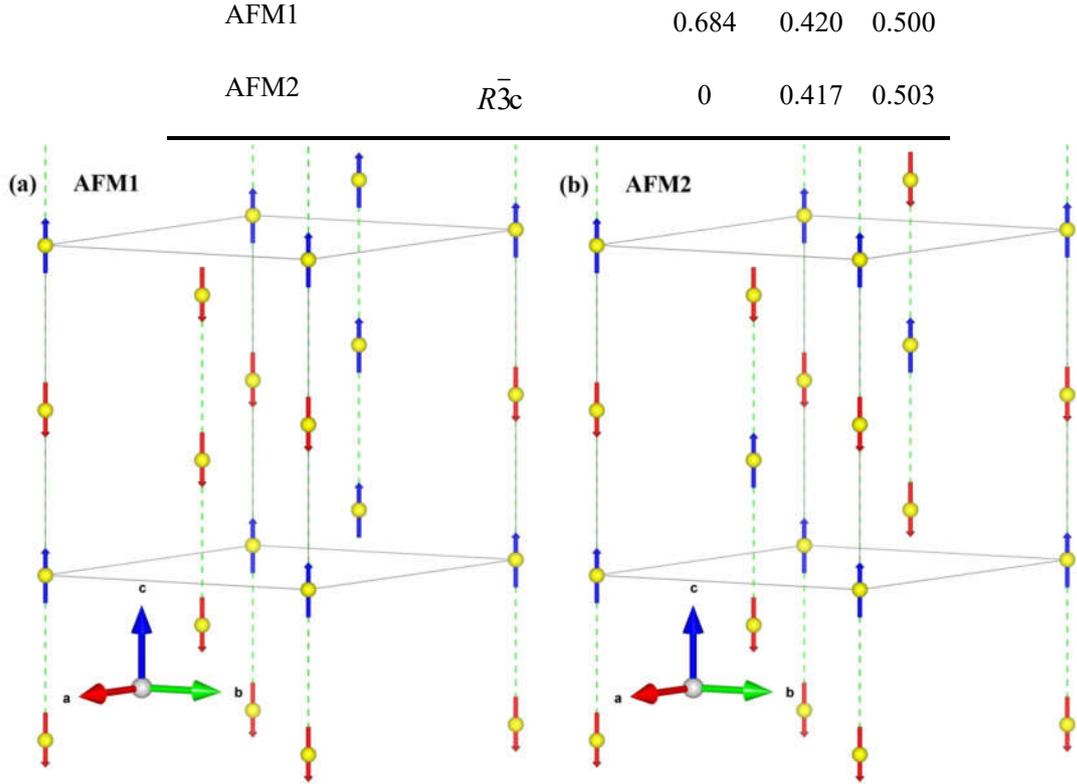

**Fig. 4** (a) Hypothetical AFM ordered spin arrangements (AFM1 state) used to extract the spin exchange constants, (b) AFM ordering (AFM2 state) corresponding to the $R\bar{3}c$ magnetic space group. Solid line indicate the hexagonal cell of $Sr_4IrO_6$, only Ir ions are shown for simplicity, and the dashed lines are guide for the eyes for the linear chains parallel to the hexagonal *c* axis. The up (down) arrows represent the magnetic moment orientations.

In order to estimate the exchange coupling constants, we artificial construct another one magnetic structure as shown in **Fig. 4** (a), denoted as AFM1 state. As shown in **TABLE II**, total energy calculations demonstrate that the AFM magnetic structure with the magnetic space group $R\bar{3}c$ (denoted as AFM2 state) is indeed energetically favorable with respect to other magnetic ordering states. At the same time, $Sr_4IrO_6$ exhibits a preferred spin orientation along the hexagonal *c* axis, which is consistent well with the symmetry requirements of the $R\bar{3}c$ magnetic space group. Such preferred orientation is primarily responsible for the magnetic ordering state in $Sr_4IrO_6$, as also found to be essential for long-range ordering in the ideal square-planar coordinated $Na_4IrO_4$ [10, 11] and double perovskite $Sr_2MIrO_6$ (M = Ca, and Mg) [53].

The calculated energy of the AFM1 state is comparable to the FM state, with energy difference less than 1 meV per hexagonal unit cell (0.14 meV/f. u.). The AFM2 state is lower in energy than the FM state by less than 1 meV per f. u. The small energy differences among these magnetic ordering states imply very weak exchange coupling interactions in $Sr_4IrO_6$. The exchange coupling constants can be derived from the corresponding energy differences of the spin ordered states by mapping the relative energies onto the classical Heisenberg model:

$$H = -\frac{1}{2}\sum_{i,j} J_{i,j} S_i \cdot S_j,$$

where negative/positive values of exchange parameter $J$ indicate AFM/FM interactions, respectively. Because of well separated $IrO_6$ octahedra and large inter-chain next nearest-neighbor (NNN) distances (about 6.94 Å) between the $Ir^{4+}$ ions, we ignore the inter-chain NNN exchange interactions. The spin exchange interactions are dominated by the intra-chain nearest-neighbor (NN) AFM exchange parameter $J_1$ (-2.76 meV), which are much stronger than the inter-chain NN FM exchange parameter $J_2$ (0.37 meV). We should note the fact that the intra-chain NN Ir-Ir distances (about 5.98 Å) are similar to the inter-chain NN distances (about 6.02Å), and a given Ir site has two intra-chain NN and six inter-chain NN $Ir^{4+}$ ions. Therefore $Sr_4IrO_6$ is three dimensional although its hexagonal crystal structure is commonly described in terms of trigonal arrangement of the $IrO_6$ chains, and the spin-spin interactions between the Ir sites cannot be one-dimensional. The calculated exchange coupling parameters explain why $Sr_4IrO_6$ undergoes a long-range AFM ordering at $T_N$ = 12 K [40]. The similarly low AFM ordering temperature also is observed in the isostructural iridate $Ca_4IrO_6$ [30, 31, 32] and 4d-based oxide $Sr_4RhO_6$ [33], reflecting the intrinsic feature of the isolated $IrO_6$ and $RhO_6$ octahedra. By comparison, the isostructural iridates $Sr_3CoIrO_6$ and $Sr_3NiIrO_6$ show a partially disordered AFM behavior due to the one-dimensional spin chains [34, 35]. Besides, the isostructural osmium oxides $Ca_3LiOsO_6$ undergoes AFM ordering at a high temperature of 117 K, implying that the spin exchange interactions are strong, and is explained as the extended superexchange interactions [54].

When the SOC interactions included, the spin magnetic moments ($M_S$) decrease from 0.430 to 0.373 $\mu_B$, and pronounced orbital moments ($M_L$) of 0.412 $\mu_B$ appear at the Ir site. When Coulomb interactions also included, $M_S$ decreases from 0.560 to 0.417 $\mu_B$, together with $M_L$ increases to 0.503 $\mu_B$ for the Ir ions. The calculated orbital moment as well as spin moment are much larger than those of $Sr_2IrO_4$ [6, 51], and are comparable to the ordered magnetic moment in the isostructural $Ca_4IrO_6$ and $Sr_4RhO_6$ [32, 33]. The big value of these orbital moments suggests that these compounds lie in the strong coupling regime of SOC interactions [53]. There are notably magnetic moment contributions from O atoms due to the hybridizations between Ir $5d$ and O $2p$ states. The calculated orbital moments are parallel to the spin moments for the $Ir^{4+}$ ions with nominal $5d^5$ electronic configurations, agreeing well with Hund's third rule. In contrast to other extensively studied iridates, the predicted total magnetic moment ($M_{total} = M_L + M_S$) in $Sr_4IrO_6$ is more close to the ideal value of 1 $\mu_B$ ($M_L = 2/3$ $\mu_B$ and $M_S = 1/3$ $\mu_B$) for the $J_{eff} = 1/2$ state [6, 30]. Normally, the $5d$ orbitals are spatially extended, so more itinerant is expected in the iridates. However, decreasing connectivity of $IrO_6$ octahedra gives rise to the enhancement of effective electronic correlations and SOC interactions, leading to the significant localized and fully polarized unmixed $J_{eff} = 1/2$ state in $Sr_4IrO_6$ [4, 9]. Therefore, due to the peculiar crystal structure, strong SOC effects and electronic correlations, in contrast to the expectation from the itinerancy of $5d$ iridates, a low-spin state with large local moment emerges in $Sr_4IrO_6$.

### D. $J_{eff} = 0$ state of the $Sr_3NaIrO_6$ and $Sr_3LiIrO_6$

As shown in **Fig. 5**, SOC interactions have profound impact on the Ir $t_{2g}$ states, but almost have no influence on the completely empty $e_g$ states of the $Sr_3NaIrO_6$ and $Sr_3LiIrO_6$. The notable feature of the band structure for these two iridates is the $t_{2g}$ states are divided into well separated $j_{eff} = 1/2$ doublet and $j_{eff} = 3/2$ quartet states [6]. The $j_{eff} = 1/2$ bands contribute to the CBM and the $j_{eff} = 3/2$ states contribute to the VBM, which are split off by an insulating gap of ~0.25 eV between them. Even without onsite Coulomb corrections, the $Sr_3NaIrO_6$ and $Sr_3LiIrO_6$ transform from the NM metallic state to NM insulating state in the presence of SOC interactions, implying the essential role of SOC interactions on the insulating nature. By introducing on-site Coulomb correlation $U$, the insulating gap are further enlarged to ~0.6 eV. The electronic structures are in line with the strong SOC limit, four electrons of the pentavalent $Ir^{5+}(5d^4)$ ions fully occupy the lower $j_{eff} = 3/2$ quadruplet, leading to NM $J_{eff} = 0$ characteristics. In the strong SOC scenario, $Sr_3NaIrO_6$ and $Sr_3LiIrO_6$ should be described as band insulator without magnetism [55, 56].

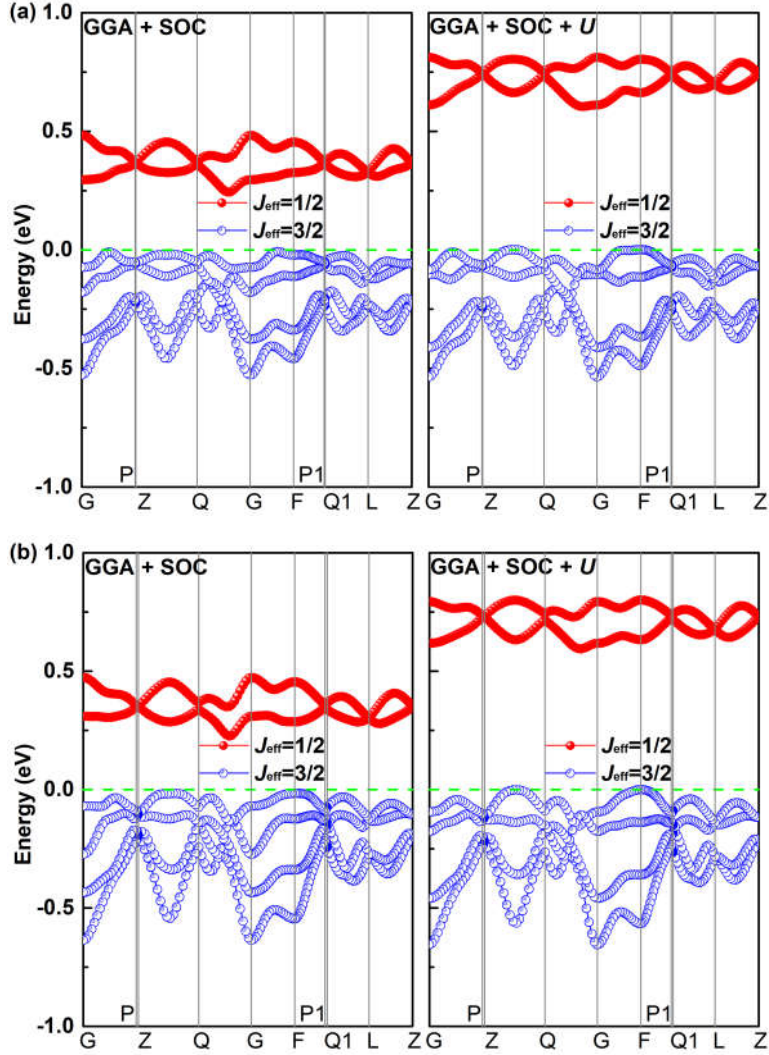

**Fig. 5** Band structures calculated within GGA + SOC (left column) and GGA + SOC + $U$ (right column): (a) $Sr_3NaIrO_6$ and (b) $Sr_3LiIrO_6$. The $j_{eff} = 1/2$ doublet are shown in red while the $j_{eff} = 3/2$ quartet states are shown in blue.

Most of previous studies have focused on the tetravalent ($Ir^{4+}$) iridates with $5d^5$ electronic configurations, whereas pentavalent ($5d^4$, $Ir^{5+}$) iridates have been rarely attracted attentions due to the expectations of NM insulating behavior. However, the delicate interplay of connectivity, SOC, Hubbard $U$, Hund's coupling and crystal field distortion gives rise to a rich magnetic phase diagram for the $d^4$ electronic configuration [56]. An absolutely NM state has been seldom observed in the $5d^4$ systems. The noncubic crystal field and the band structure effect have been proposed to result in the breakdown of the $J_{eff} = 0$ NM states in double-perovskite iridates with pentavalent ($5d^4$) $Ir^{5+}$ ions [41, 57]. The electronic correlations and SOC interactions closely link with the connectivity of the $IrO_6$ octahedra [4, 9, 57]. The almost not-distorted $IrO_6$ octahedra are well separated with each other, leading to disconnected geometry and large Ir-Ir separation, thereby bringing the situation closer to the atomic limit in hexagonal $Sr_3NaIrO_6$ and $Sr_3LiIrO_6$. The effective electron correlations are increasing along with the decreasing of the connectivity of the $IrO_6$ octahedra, accompanied by further enhancement of the SOC effect [4, 57]. In contrast to the three-dimensional connectivity in double-perovskite structure and the highly distorted $IrO_6$ octahedra in $NaIrO_3$, the local $IrO_6$ octahedra in $Sr_3NaIrO_6$ and $Sr_3LiIrO_6$ reside much more close to the ideal cubic crystal-field limit, and achieve the $J_{eff} = 0$ NM state by reducing the connectivity of the $IrO_6$ octahedra.

### E. Robust of the insulating $J_{eff}$ states in $Sr_3MIrO_6$

To shed more light on the nature of the $J_{eff}$ states, we perform non-spin polarized calculations for the NM state of $Sr_3MIrO_6$ using the projection-embedding implementation [58] on top of the WIEN2K package [59]. The structures relaxed by VASP are used to do self-consistent calculation and converge the charge density in WIEN2K by the local density approximation (LDA) method. The partial density of states (pDOS) is given by the imaginary part of the Green's function ($-1/\pi \operatorname{Im} G(E)$), which are presented in **Fig. 6**. The most striking features of $-1/\pi \operatorname{Im} G(E)$ are the almost unmixed characteristics of the $j_{eff} = 1/2$ states, which is distinctly separated from the fully-occupied quartet $j_{eff} = 3/2$ states. In contrast, the localized structural distortions of the $IrO_6$ octahedra in other iridates often lead to a mixture of the $j_{eff} = 1/2$ states with the $j_{eff} = 3/2$ states [4]. Similar to the isostructural iridate $Ca_4IrO_6$ [32], and the fluorido-iridates molecular [49], as well as the hexafluoro iridates $Rb_2IrF_6$ [50], the hexagonal iridates $Sr_3MIrO_6$ provide nearly ideal octahedral crystal-field environment to realize a pure $J_{eff} = 1/2$ state.

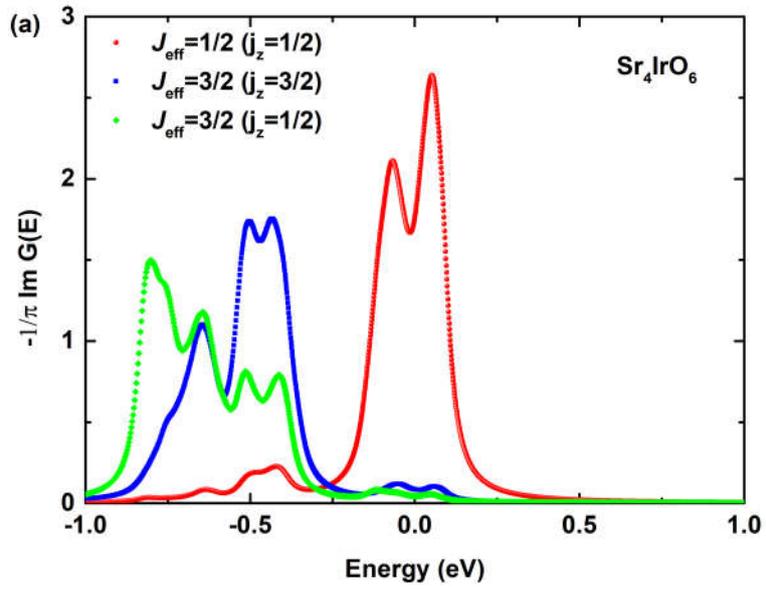

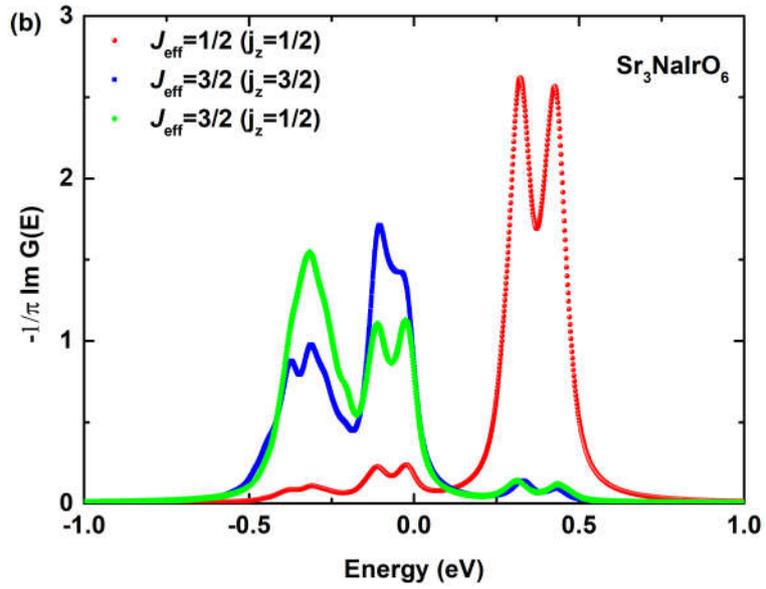

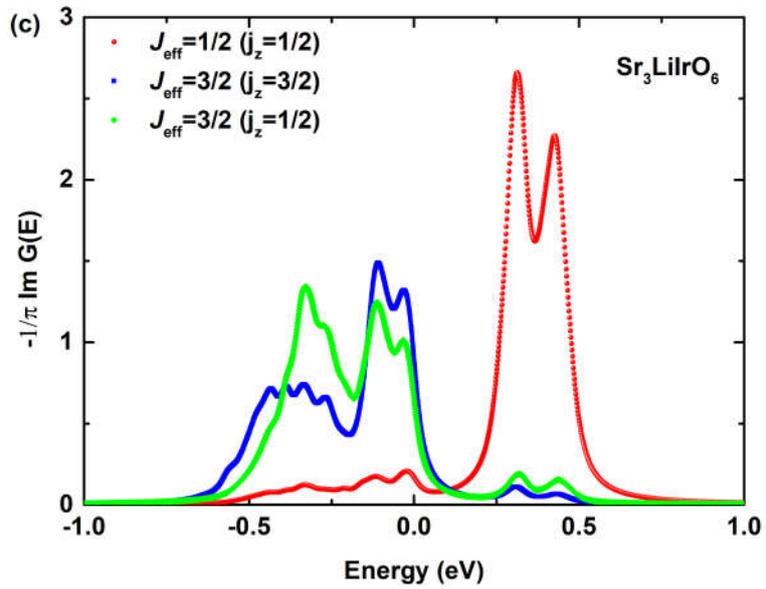

**Fig. 6** Imaginary part of the Green's functions $-1/\pi \, \text{Im} \, G(E)$ for the NM state of $Sr_3MIrO_6$: (a) $Sr_4IrO_6$, (b) $Sr_3NaIrO_6$, (c) $Sr_3LiIrO_6$.

For $Sr_4IrO_6$, the unmixed $J_{eff} = 1/2$ states cross the Fermi level, resulting in a metallic electronic structure. The calculated pDOS shows relatively sharp peaks at the Fermi levels, indicating that the NM metallic electronic structure is unstable with respect to long-range magnetic ordering in $Sr_4IrO_6$. As shown in **Fig. 3**, the introduction of AFM ordering indeed splits off the $j_{eff} = 1/2$ bands, resulting in an insulating state with a gap around 0.2 eV. For the cases of $Sr_3NaIrO_6$ and $Sr_3LiIrO_6$ with $d^4$ electronic configurations, the $j_{eff} = 1/2$ states are shifted to the upper region of the conduction bands, establishing the NM $J_{eff} = 0$ state and the insulating gap. For these hexagonal iridates, insulating gap can open up when SOC interactions are took into account even without Coulomb corrections. The spatially isolated octahedra with tiny distortions associated with SOC interactions are crucial in producing robust insulating $J_{eff}$ states in these hexagonal iridates.

In the high-symmetry phase of the transition metal compounds with cubic crystal-field, the three degenerate $t_{2g}$ states often give rise to metallic electronic structures in the case of partially filled shell. It is hard to open the gap due to the degenerate energy levels. Two scenarios, magnetically driven (Slater transition) and Coulomb repulsion driven (Mott-Hubbard transition) metal-insulator transition (MIT) have been proposed to describe the insulating nature in transition metal compounds [60]. Usually, the structural distortions or the Jahn-Teller effect break the symmetry and remove the degeneracy of the electronic states, consequently the gap is opened with the help of the Coulomb repulsion [61]. However, in these iridates compounds, this degeneracy inhibits the rise of an insulating phase even in the case of large value of the Coulomb repulsion. The role of the symmetry breaking of the $t_{2g}$ manifold degeneracy is played by the SOC. The large SOC produces the splitting between the $j_{eff} = 1/2$ and the $j_{eff} = 3/2$ states. Finally, the Coulomb interaction enlarges the gap opened by the SOC.

## IV. CONCLUSIONS

In conclusion, the hexagonal $Sr_3MIrO_6$ serves as a canonical model system to investigate the underlying physical properties that arises from the novel $J_{eff}$ state. The strong SOC interactions play a crucial role in realizing the robust insulating $J_{eff}$ states in the hexagonal iridates $Sr_3MIrO_6$, as shown by our comprehensive DFT calculations joint with Green's functions analysis. The decreasing connectivity of $IrO_6$ octahedra gives rise to the increasing of effective electronic correlations and SOC interactions, leading to low-spin $J_{eff}$ = 1/2 states with large magnetic moments and weak spin exchange interactions for the $Ir^{4+}$ ($5d^5$) ions in $Sr_4IrO_6$, and NM singlet $J_{eff}$ = 0 states without magnetic moments for the $Ir^{5+}$ ($5d^4$) ions in $Sr_3NaIrO_6$ and $Sr_3LiIrO_6$. We hope our theoretical simulations will stimulate experimental works aimed at detailed magnetic properties measurements and characterizations, to further understand the magnetic ground state and exploit other physical properties in these sparsely-studied hexagonal iridates.

## ACKNOWLEDGMENTS

We acknowledge Silvia Picozzi, Giorgio Sangiovanni and Mario Cuoco for useful discussions. X. M. was sponsored by the Guangxi Natural Science Foundation and the Scientific Research Foundation of Guilin University of Technology (No. GUTQDJJ2017105). C. A. was supported by CNR-SPIN via the Seed Project CAMEO. High performance computational resources provided by National Supercomputer Center on TianHe-2 in LvLiang of China are also gratefully acknowledged.